\documentclass[aps,prb,twocolumn,superscriptaddress,showpacs,floatfix]{revtex4}
\usepackage{times}
\usepackage{graphicx}
\usepackage{epsfig}
\usepackage{amsmath}
\usepackage{amssymb}
\usepackage{natbib}
\usepackage{color}

\begin{document}

\title{Modeling Spin Dynamics in the Singlet Ground State Garnet Ho$_3$Ga$_5$O$_{12}$}

\author{Joseph~A.~M.~Paddison}
\affiliation{Churchill College, University of Cambridge, Storey's Way, Cambridge CB3 0DS, UK}
\affiliation{Cavendish Laboratory, University of Cambridge, JJ Thomson Ave, Cambridge CB3 0HE, UK}
\affiliation{School of Physics, Georgia Institute of Technology, 837 State Street, Atlanta, Georgia 30332, USA}

\author{Paromita Mukherjee}
\affiliation{Cavendish Laboratory, University of Cambridge, JJ Thomson Ave, Cambridge CB3 0HE, UK}

\author{Xiaojian Bai}
\affiliation{School of Physics, Georgia Institute of Technology, 837 State Street, Atlanta, Georgia 30332, USA}

\author{Zhiling Dun}
\affiliation{School of Physics, Georgia Institute of Technology, 837 State Street, Atlanta, Georgia 30332, USA}

\author{Christopher R. Wiebe}
\affiliation{Department of Chemistry, University of Winnipeg, 515 Portage Ave, Winnipeg, MB, R3B 2E9, Canada}
\affiliation{Department of Physics, Florida State University, Tallahassee, Florida 32306-3016, USA}
\affiliation{National High Magnetic Field Laboratory, Florida State University, Tallahassee, Florida 32310, USA}

\author{Haidong Zhou}
\affiliation{Department of Physics and Astronomy, University of Tennessee, Knoxville, Tennessee 37996, USA}

\author{Jason S. Gardner}
\affiliation{Songshan Lake Materials Laboratory, Dongguan, Guangdong 523808, China}

\author{Martin Mourigal}
\affiliation{School of Physics, Georgia Institute of Technology, 837 State Street, Atlanta, Georgia 30332, USA}

\author{Si\^{a}n~E.~Dutton}
\affiliation{Cavendish Laboratory, University of Cambridge, JJ Thomson Ave, Cambridge CB3 0HE, UK}

\date{\today}

\begin{abstract}
Materials containing non-Kramers magnetic ions can show unusual quantum excitations because of the exact mapping of the two singlet crystal-field ground state to a quantum model of Ising spins in a transverse magnetic field. Here, we model the magnetic excitation spectrum of garnet-structured Ho$_3$Ga$_5$O$_{12}$, which has a two-singlet crystal-field ground state. We use a reaction-field approximation to explain published inelastic neutron-scattering data [Zhou \emph{et al.}, \emph{Phys. Rev. B} \textbf{78}, 140406(R) (2008)] using a three-parameter model containing the magnetic dipolar interaction, the two-singlet crystal-field splitting, and the nuclear hyperfine coupling. Our study clarifies the magnetic Hamiltonian of Ho$_3$Ga$_5$O$_{12}$, reveals that the nuclear hyperfine interaction drives magnetic ordering in this system, and provides a framework for quantitative analysis of magnetic excitation spectra of materials with singlet crystal-field ground states.
\end{abstract}

\maketitle

\section{Introduction}

Materials that realize simple models of quantum magnetism are of fundamental interest because they allow current theories to be tested experimentally.\cite{Keimer_2017}
The excitation spectrum of a magnetic material---measurable by inelastic neutron-scattering experiments---directly probes the spin dynamics of the system, and is often sensitive to the properties of the underlying quantum model.\cite{Marshall_1968}
Arguably the two simplest examples of magnetic excitations are spin waves (magnons) and crystal-field excitations. The former describe the collective magnetic excitations of ordered magnetic states in terms of the precession of spins about their average directions. The latter describe single-ion excitations: for a rare-earth metal ion with angular-momentum quantum number $J$, they correspond to transitions between the $2J+1$ levels that are split by the crystalline electric field (CEF).\cite{Jensen_1991} In many magnetic materials, the energy scale of the relevant CEF transitions is either much larger than the magnetic interactions (``CEF limit"), or much smaller than the magnetic interactions (``spin-wave limit"). In the CEF limit, the system typically shows no collective magnetic ordering and its excitations are CEF.\cite{Zinkin_1996} In the spin-wave limit, the system typically orders magnetically, and its low-energy excitations are spin waves.\cite{Fulde_1985}

\begin{figure}
\begin{center}
\includegraphics[scale=1.]{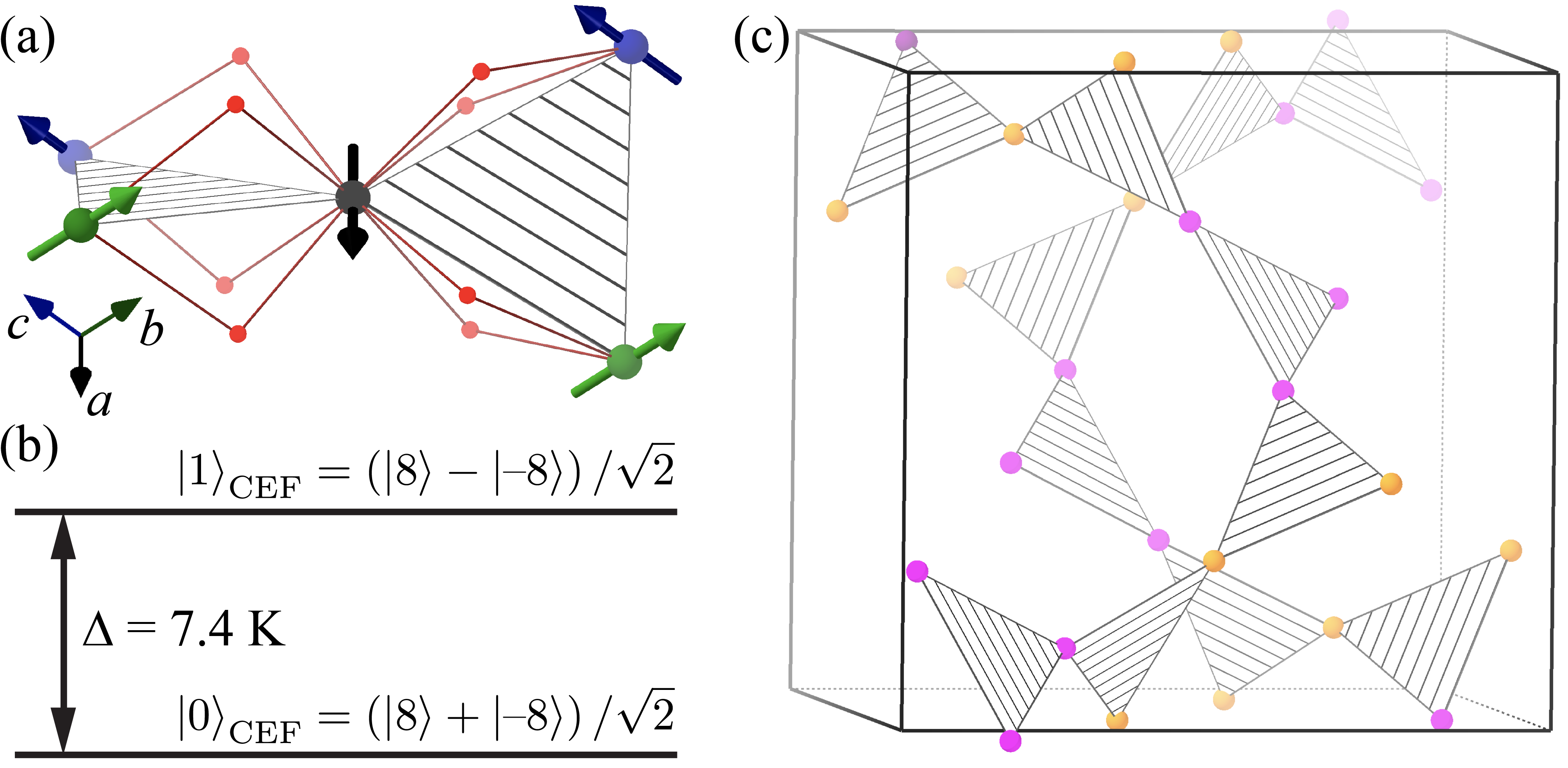}
\end{center}
\caption{\label{fig1}(a) Local Ho$^{3+}$ environment, showing the four Ho$^{3+}$ neighbors (large circles) and eight O$^{2-}$ neighbors (small circles) of a central Ho$^{3+}$ ion. The local $\mathbf{z}$ quantization axes are shown as black, blue, and green arrows. Each quantization axis is a $C_2$ axis of point symmetry and a global $\langle 100\rangle$ direction. Nearest-neighbor spins are orthogonal. (b) Low-energy crystalline electric field (CEF) levels of Ho$^{3+}$  ions in Ho$_3$Ga$_5$O$_{12}$, which comprise two singlets separated by an energy gap $\Delta= 7.4$\,K. (c) Partial crystal structure of Ho$_3$Ga$_5$O$_{12}$ showing corner-sharing triangles of Ho$^{3+}$ ions (circles) colored according to the Ho$^{3+}$ spin orientations in the magnetically-ordered state: gold if the spin is parallel to its quantization axis, and purple if it is antiparallel.}\end{figure}

Materials in which magnetic interactions and CEF are comparable in energy can show unusual quantum behavior.\cite{Le_2011,Choi_2012} The simplest example is a CEF that yields only two singlets at thermally-accessible energies.\cite{Wang_1968} This situation can occur for non-Kramers ions; i.e., those with integer $J$.
The two singlets are separated by an energy $\Delta$, and angular momenta are coupled by a pairwise interaction $\mathcal{K}$. In the CEF limit ($\mathcal{K}=0$), the excitation spectrum measured by neutron scattering consists of a non-dispersive CEF mode at an energy transfer of $\Delta$. 
If the magnitude of $\mathcal{K}$ is increased, the paramagnetic excitation spectrum acquires dispersion, and begins to soften at an incipient ordering wavevector. A soft-mode transition to a magnetically-ordered state only occurs if the ratio $\mathcal{K}/\Delta$ exceeds a critical value; otherwise, the mode softening remains partial and the system remains in a correlated paramagnetic phase to zero temperature.\cite{Wang_1968,Brout_1966}

The interacting two-singlet model with $\mathcal{K}\sim\Delta$ is of fundamental interest because it maps to a canonical model of quantum mechanics---an effective spin-$1/2$ Ising model in a transverse magnetic field.\cite{Stinchcombe_1973,Suzuki_2012} This model has been applied to diverse physical phenomena in which quantum tunneling competes with pairwise interactions, such as the soft-mode transitions in order-disorder ferroelectrics.\cite{Brout_1966}
Remarkably, recent theoretical work has shown that the combination of two-singlet spin dynamics with frustrated magnetic interactions can yield a quantum spin-liquid phase on the pyrochlore lattice.\cite{Savary_2017} It has been proposed that such a state may be realized experimentally in Pr$^{3+}$-based pyrochlore magnets\cite{Zhou_2008a,Sibille_2016}---in which structural disorder can generate random CEF splittings (transverse fields)\cite{Wen_2017,Martin_2017}---and may also be relevant to the low-temperature behavior of Tb$_2$Ti$_2$O$_7$.\cite{Bonville_2011,Petit_2012,Gardner_1999,Gingras_2000} The potential for exotic quantum behavior in interacting two-singlet systems demonstrates the need to identify real materials of this type and to benchmark current theories against high-quality neutron-scattering data.

The cubic rare-earth garnet Ho$_3$Ga$_5$O$_{12}$ is a candidate material to realize the interacting two-singlet model. In this system, the crystal-field Hamiltonian of the magnetic Ho$^{3+}$ ions has been investigated by multiple experimental techniques, including magnetic susceptibility,\cite{Cooke_1967} neutron diffraction,\cite{Hammann_1977,Hammann_1977a} inelastic neutron scattering,\cite{Reid_1991,Zhou_2008} optical spectroscopy,\cite{Johnson_1970} and specific heat.\cite{Onn_1967}
In Ho$_3$Ga$_5$O$_{12}$, non-Kramers Ho$^{3+}$ ions with $J=8$  occupy a site with low point symmetry ($D_2$), as shown in Fig.~\ref{fig1}(a). Consequently, all crystal-field levels in Ho$_3$Ga$_5$O$_{12}$ are singlets.\cite{Walter_1984} Neutron spectroscopy\cite{Reid_1991,Nekvasil_1979}  and specific-heat\cite{Onn_1967} measurements show that the two lowest-energy singlets are separated by $\Delta = 7.4$\,K, as shown in Fig.~\ref{fig1}(b). Because the third CEF level is at approximately $50$\,K,\cite{Reid_1991} at low temperatures ($T\ll 50$\,K) only the lowest-energy two singlets are thermally populated and the system approximates a two-singlet CEF ground state. The CEF generates an Ising anisotropy of the magnetic moments, which lie parallel or antiparallel to the local $\mathbf{z}\in\langle 100\rangle$ axes shown in Fig.~\ref{fig1}(b).\cite{Nekvasil_1979} The approximate magnitude of the magnetic dipolar interaction $4DJ^2 = 4.68$\,K at the nearest-neighbor distance is only moderately smaller than $\Delta$; hence, the system is intermediate between spin-wave and CEF limits. 

Whereas early reports\cite{Cooke_1967,Hammann_1977,Hammann_1977a,Reid_1991,Johnson_1970,Onn_1967} focused on the single-ion properties of Ho$_3$Ga$_5$O$_{12}$, recent studies\cite{Zhou_2008,Mukherjee_2017} have focused on the potential for collective phenomena induced by geometrical frustration. This possibility occurs because the Ho$^{3+}$ ions in Ho$_3$Ga$_5$O$_{12}$ (space group $Ia\bar{3}d$) occupy two interpenetrating ``hyperkagome" networks of corner-sharing triangles [Fig.~\ref{fig1}(c)], which would be frustrated for antiferromagnetic nearest-neighbor interactions.\cite{Yoshioka_2004,Petrenko_2000,Yavorskii_2006,Paddison_2015} Experimentally, Ho$_3$Ga$_5$O$_{12}$ orders magnetically at a $T_\mathrm{N}$ between $0.15$\,K and $0.3$\,K,\cite{Hammann_1977,Hammann_1977a,Zhou_2008} which is significantly smaller than the energy scale of dipolar interactions. However, two further effects need to be considered in Ho$_3$Ga$_5$O$_{12}$.
The first is the two-singlet CEF discussed above, which is expected to suppress $T_\mathrm{N}$.\cite{Wang_1968} The second is the hyperfine interaction between electronic and nuclear spins, which is significant because of the large values of both electronic and nuclear spin quantum numbers for Ho$^{3+}$ ($J= 8$ and $I=7/2$, respectively), and is expected to enhance $T_\mathrm{N}$.\cite{Hammann_1973,Andres_1973,Murao_1971,Murao_1975} The interplay of spin-spin interactions, hyperfine interactions, and the two-singlet CEF in Ho$_3$Ga$_5$O$_{12}$ remains an open question.

Here, we present a modeling study of previously-published inelastic neutron-scattering data on Ho$_3$Ga$_5$O$_{12}$ (from Zhou \emph{et al.}, Ref.~\onlinecite{Zhou_2008}). 
Our analysis reveals four key results. First, we explain the experimental data using a magnetic Hamiltonian that contains three terms: the crystal-field splitting $\Delta$, the long-range dipolar coupling $D$, and the hyperfine coupling $A$, where the values of all three parameters are fixed from first principles or from previous experiments. Second, we find that geometrical frustration does not play a significant role in the behavior of Ho$_3$Ga$_5$O$_{12}$. Third, we find that the nuclear hyperfine interaction actually drives the transition to a magnetically-ordered state. 
Fourth, we show that a self-consistent modification of mean-field/RPA theory---the reaction-field approximation\cite{Santos_1980,Brout_1967}---yields a significantly better description of the experimental data than standard RPA. Our results have implications for the quantitative modeling of topical frustrated materials in which interactions and crystal-field transitions have similar energy scales, such as Pr$^{3+}$-based pyrochlores,\cite{Zhou_2008a,Wen_2017,Martin_2017} osmate pyrochlores,\cite{Zhao_2016} spin-chain SrHo$_2$O$_4$,\cite{Young_2013} spinel NiRh$_2$O$_4$,\cite{Chamorro_2018} and tripod kagome systems.\cite{Dun_2016,Dun_2017,Paddison_2016, Ding_2018} 

Our paper is structured as follows. In Section \ref{sec:model}, we introduce the magnetic Hamiltonian we use to model Ho$_3$Ga$_5$O$_{12}$ and the reaction-field theory we use to calculate the magnetic excitation spectrum. In Section \ref{sec:results}, we compare our model calculations with previously-published single-crystal neutron-scattering data,\cite{Zhou_2008} and discuss the implications for the physics of Ho$_3$Ga$_5$O$_{12}$.  We conclude in Section \ref{sec:conclusions} with a discussion of the applications of our study.

\section{Model}\label{sec:model}

\subsection{Magnetic Hamiltonian}

We consider the following magnetic Hamiltonian for Ho$_3$Ga$_5$O$_{12}$,
\begin{eqnarray}
\mathcal{H} = & = & \sum_{i,\mathbf{r}}\left[\mathcal{H}_{i}^{\mathrm{CEF}}(\mathbf{r})+\mathcal{H}_{i}^{\mathrm{HF}}(\mathbf{r})\right]\nonumber \\
 &  & -\frac{1}{2}\sum_{i,j}\sum_{\mathbf{r},\mathbf{r}^{\prime}}\mathcal{K}_{ij}(\mathbf{r}^{\prime}-\mathbf{r})J_{i}^{z}(\mathbf{r})J_{j}^{z}(\mathbf{r}^{\prime}),\label{eq:hamiltonian}
\end{eqnarray}
where $\mathcal{H}_{i}^{\mathrm{CEF}}$ is the crystal-field Hamiltonian for atom $i$ in the unit cell located at lattice vector $\mathbf{r}$,  $\mathcal{H}_{i}^{\mathrm{HF}}$ is the nuclear hyperfine Hamiltonian, $\mathcal{K}_{ij}$ is the pairwise interaction between angular momenta $i$ and $j$ at lattice vectors $\mathbf{r}$ and $\mathbf{r}^{\prime}$, respectively, and $J^z_i$ is the operator for the $z$-component of electronic angular momentum, where the local $\mathbf{z}$ axes of quantization are shown in Fig.~\ref{fig1}(a). The  low-energy crystal-field Hamiltonian $H_{\mathrm{CEF}}$ comprises two singlets that are separated by $\Delta = 7.4$\,K and have wavefunctions
\begin{eqnarray}
|{0}\rangle \approx \frac{1}{\sqrt{2}} (|{8}\rangle+|{-8}\rangle), \nonumber\\
|{1}\rangle \approx \frac{1}{\sqrt{2}} (|{8}\rangle-|{-8}\rangle), \label{eq:singlets}
\end{eqnarray} 
where $|\pm8\rangle$ is shorthand for $|J=8,J^{z}=\pm8\rangle$.\cite{Reid_1991,Nekvasil_1979} Although both singlets are individually nonmagnetic (i.e., $\langle 0 |{J}^z |0\rangle = \langle 1|{J}_z |1\rangle =0$), the nonzero matrix element $m=|\langle 0 |{J}^z |1\rangle|\approx8$ between the two singlets gives rise to a paramagnetic moment.
The hyperfine Hamiltonian
\begin{equation}
\mathcal{H}_{i}^{\mathrm{HF}}(\mathbf{r})= AJ_{i}^{z}(\mathbf{r})I_{i}^{z}(\mathbf{r}),
\end{equation}
where $A=0.039$\,K is the nuclear hyperfine coupling constant for Ho,\cite{Krusius_1969,Ramirez_1994} $I^z$ is the $z$-component of nuclear angular momentum, and $I=7/2$ for $^{165}$Ho. The pairwise interaction $\mathcal{K}_{ij}$ may, in principle, include contributions from both exchange interactions and the long-range dipolar interaction. However, because the local Ising axes enforce that any central spin is orthogonal to its nearest neighbors [Fig.~\ref{fig1}(a)], the isotropic contribution to the nearest-neighbor exchange is zero. As a first approximation, therefore, we consider only the long-range dipolar interaction
\begin{equation}
\mathcal{K}_{ij}^{\mathrm{dip}}(\mathbf{r}^{\prime}-\mathbf{r})=D|\mathbf{r}_{\mathrm{nn}}|^{3}\left[\frac{3(\hat{\mathbf{z}}_{i}\cdot\hat{\mathbf{r}}_{ij})(\hat{\mathbf{z}}_{j}\cdot\hat{\mathbf{r}}_{ij})-\hat{\mathbf{z}}_{i}\cdot\hat{\mathbf{z}}_{j}}{|\mathbf{r}_{ij}|^{3}}\right],\label{eq:dipolar}
\end{equation}
between angular momenta separated by a vector $\mathbf{r}_{ij}=\mathbf{r}^\prime+\mathbf{R}_{j}-\mathbf{r}-\mathbf{R}_{i}$, where $\hat{\mathbf{r}}_{ij}=\mathbf{r}_{ij}/|\mathbf{r}_{ij}|$, and $D= \mu_{0}(g_{J}\mu_{\mathrm{B}})^{2}/{4\pi}{k_{\mathrm{B}}|\mathbf{r}_{\mathrm{nn}}|^{3}}=0.0183$\,K is determined by the nearest-neighbor distance $|\mathbf{r}_{\mathrm{nn}}|=3.76$\,\AA,~and the Lande $g$-factor $g_{J}=5/4$ for Ho$^{3+}$.

\subsection{RPA and reaction-field approximation}

We model the magnetic Hamiltonian [Eq.~\eqref{eq:hamiltonian}] using the random-phase approximation (RPA) with a reaction-field term intended to account for local magnetic correlations.\cite{Brout_1967,Logan_1995} 
The reaction-field method has previously been derived for the Ising model in a transverse field for systems where there is a single magnetic ion in the primitive unit cell, $N=1$.\cite{Santos_1980} Here, we extend the approach to calculate the paramagnetic ($T>T_\mathrm{N}$) excitation spectra for systems with $N>1$, as is required for the Ho$^{3+}$ ions in Ho$_3$Ga$_5$O$_{12}$ ($N=12$). The equations given below are general and can be applied directly to other lattices. Throughout, we work in natural units with $\hbar=k_\mathrm{B}=1$.

We consider the magnetization induced by
applying a time-dependent and site-dependent field $h_{i}^{z}(\mathbf{Q},t)$ to a paramagnetic system, which leads
to an effective field given by
\begin{equation}
h_{i,\mathrm{eff}}^{z}(\mathbf{Q},t)=h_{i}^{z}(\mathbf{Q},t)+\sum_{j}\left[\mathcal{K}_{ij}(\mathbf{Q})-\lambda\right] J_{j}^{z}(\mathbf{Q},t),\label{eq:effective_field}
\end{equation}
where $J_{i}^{z}(\mathbf{Q},t)=\sum_{\mathbf{r}}J_{i}^{z}(\mathbf{r},t)\exp(\mathrm{i}\mathbf{Q}\cdot\mathbf{r})$ is the Fourier transform of the angular momentum of atom $i$ in the primitive cell, and $\mathcal{K}_{ij}(\mathbf{Q})$ is the Fourier transform of the magnetic interactions between atoms $i$ and $j$. The long-ranged nature of the magnetic dipolar interaction is handled using Ewald summation techniques.\cite{Enjalran_2004} The reaction field $\lambda$ is motivated by the fact that a spin is not affected by the field due to its own orientation, and will be determined below.
For a system containing $N$ magnetic atoms in its primitive unit cell, the dynamical magnetic susceptibility is calculated by decomposing $J_{i}^{z}(\mathbf{Q},t)$ into its $N$ normal modes,\cite{Enjalran_2004}
\begin{equation}
J_{i}^{z}(\mathbf{Q},t) =\sum_{\mu}m_{\mu}(\mathbf{Q},t)U_{i\mu}(\mathbf{Q})\label{eq:magn},
\end{equation}
where $m_{\mu}(\mathbf{Q},t)$ is the amplitude of the mode $\mu$. An analogous mode decomposition is performed for the applied field $h_i^{z}(\mathbf{Q},t) $. The eigenenergies
$\lambda_{\mu}$ and eigenvector components $U_{i\mu}$ are given at each $\mathbf{Q}$ as the solutions of
\begin{equation}
\lambda_{\mu}(\mathbf{Q})U_{i\mu}(\mathbf{Q})=\sum_{j}\mathcal{K}_{ij}(\mathbf{Q})U_{j\mu}(\mathbf{Q}).\label{eq:eigenvalue}
\end{equation}
From Eqs.~(\ref{eq:effective_field})--(\ref{eq:eigenvalue}), the susceptibility $\chi_{\mu}=J^z_{\mu}/h^z_{\mu}$ for each mode is given by
\begin{equation}
\chi_{\mu}(\mathbf{Q},\omega)=\frac{\chi_{0}(\omega)}{1-\chi_{0}(\omega)[\lambda_{\mu}(\mathbf{Q})-\lambda]},\label{eq:rpa}
\end{equation}
which is identical to the RPA expression\cite{Brout_1966} except for the appearance of $\lambda$. The noninteracting (single-ion) susceptibility is the sum of an inelastic contribution from nondegenerate states and a quasielastic Curie-law contribution law from degenerate states,\cite{Fulde_1985}
\begin{eqnarray}
\chi_{0}&=&\sum_{\omega_{ij}\neq0}\frac{|m_{ij}|^{2}\omega_{ij}(n_{i}-n_{j})}{\omega_{ij}^{2}-\omega^{2}-\mathrm{i}\omega\Gamma}+\delta_{\omega0}\sum_{\omega_{ij}=0}\frac{|m_{ij}|^{2}n_{i}}{T}\\
&=& \chi_{0}^{\mathrm{in}}(\omega)+\delta_{\omega0}\chi_{0}^{\mathrm{el}},
\label{eq:chi_singleion}
\end{eqnarray}
where $n_{i}=\exp(-\omega_{i}/T)/\sum_{j}\exp(-\omega_{j}/T)$ is a thermal population factor, $\omega_{ij}=\omega_{j}-\omega_{i}$ is an energy difference between states, and $\Gamma$ is a small, positive relaxation rate. We obtain $\chi_{0}(\omega)$ by exact diagonalization of the single-ion Hamiltonian for Ho$_3$Ga$_5$O$_{12}$ in the $16\times16$-dimensional space formed by $2I+1$ nuclear-spin states and two electronic-spin states; the $\omega_{ij}$ and transition dipole matrix elements $m_{ij}$ are derived in Appendix A. The hyperfine interaction splits the two electronic singlets into a total of 8 doublets; this doublet degeneracy occurs because the nuclear spins ($I=7/2$) possess Kramers degeneracy.
From Eqs.~(\ref{eq:rpa})--(\ref{eq:chi_singleion}), the inelastic and quasi-elastic contributions to the interacting susceptibility are given by
\begin{equation}
\chi^{\mathrm{in}}_{\mu}(\mathbf{Q},\omega)=\frac{\chi_{0}^{\mathrm{in}}(\omega)}{1-\chi_{0}^{\mathrm{in}}(\omega)\left[\lambda_{\mu}(\mathbf{Q})-\lambda\right]}
\label{eq:chi_inel}
\end{equation}
and
\begin{equation}
\chi^{\mathrm{el}}_{\mu}(\mathbf{Q})=\frac{\chi_{0}^{\mathrm{el}}}{1-\left[\chi_{0}^{\mathrm{el}}+\chi_{0}^{\mathrm{in}}(0)\right]\left[\lambda_{\mu}(\mathbf{Q})-\lambda\right]},
\label{eq:chi_el}
\end{equation}
respectively.

Following Ref.~\onlinecite{Santos_1980}, the reaction field $\lambda$ is determined by enforcing the sum rule on the magnitude of the electronic angular momentum,
\begin{equation}
\frac{1}{NN_{\mathbf{q}}}\sum_{i,\mathbf{q}}\langle J_{i}^{z}(\mathbf{q},0)J_{i}^{z}(-\mathbf{q},0)\rangle =m^{2},\label{eq:sum_rule}
\end{equation}
where the sum is over wavevectors $\mathbf{q}$ in the first Brillouin
zone and spins $i$ in the primitive cell, and $m=8$ is the electronic angular momentum of the two-singlet ground state in Ho$_3$Ga$_5$O$_{12}$.
We express this sum rule in terms of the dynamical susceptibility by combining Eqs.~\eqref{eq:magn} and \eqref{eq:sum_rule} with the fluctuation-dissipation theorem to obtain the self-consistency
equation \cite{Santos_1980} 
\begin{multline}
m^{2} = \frac{1}{NN_{\mathbf{q}}}\sum_{\mu,\mathbf{q}} \big[ T\chi^{\mathrm{el}}_{\mu}(\mathbf{q}) \\
+ \frac{1}{\pi} \int_{-\infty}^{\infty}[n(\omega)+1]{\mathrm{Im}[\chi^{\mathrm{in}}_{\mu}(\mathbf{q},\omega)]}\mathrm{d}\omega \big] ,\label{eq:onsager}
\end{multline}
where $n(\omega)=[\exp(\omega/T)-1]^{-1}$ is the Bose population factor. Eq.~\eqref{eq:onsager} is solved numerically for $\lambda$ at each temperature.

\subsection{Neutron-scattering intensity: general expression}
The magnetic neutron-scattering intensity is given by $I(\mathbf{Q},\omega)=C[g_{J}f(|\mathbf{Q}|)]^2S(\mathbf{Q},\omega)$, where $f(|\mathbf{Q}|)$ is the magnetic form factor,\cite{Brown_2004} $C=(\gamma_{\mathrm{n}}r_{0}/2)^2=0.07265$\,barn is a constant, and $S(\mathbf{Q},\omega)$ is the magnetic neutron-scattering function given by
\begin{multline}
S(\mathbf{Q},\omega)  =\frac{1}{2\pi N}\sum_{i,j}\int_{-\infty}^{\infty}\left\langle \mathbf{J}_{i}^{\perp}(-\mathbf{Q},0)\cdot\mathbf{J}_{j}^{\perp}(\mathbf{Q},t)\right\rangle \\
  \times\exp\left[\mathrm{i}\mathbf{Q}\cdot(\mathbf{r}_{j}-\mathbf{r}_{i})-\mathrm{i}\omega t\right]\mathrm{d}t,\label{eq:intensity}%\\
\end{multline}
where $\mathbf{J}^{\perp}_{i}=J^{z}_{i}\mathbf{z}_{i}^{\perp}$, with $\mathbf{z}_{i}^{\perp}=\hat{\mathbf{z}_{i}}-(\hat{\mathbf{z}_{i}}\cdot\hat{\mathbf{Q}})\hat{\mathbf{Q}}$ the component of the local Ising axis perpendicular to $\mathbf{Q}$.\cite{Lovesey_1987}
From Eqs.~\eqref{eq:effective_field}--\eqref{eq:onsager}, we obtain $S(\mathbf{Q},\omega)$ in the paramagnetic phase as the sum of inelastic and quasi-elastic contributions,
\begin{equation}
S_\mathrm{in}(\mathbf{Q},\omega) =\frac{n(\omega)+1}{\pi N}\sum_{\mu}\left|\mathbf{F}_{\mu}^{\perp}(\mathbf{Q})\right|^{2}\mathrm{Im}[\chi_{\mu}^{\mathrm{in}}(\mathbf{Q},\omega)]\label{eq:s_qe_inel}
\end{equation}
and
\begin{equation}
S_\mathrm{el}(\mathbf{Q},\omega) =\frac{T}{N}\sum_{\mu}\left|\mathbf{F}_{\mu}^{\perp}(\mathbf{Q})\right|^{2}\chi_{\mu}^{\mathrm{el}}(\mathbf{Q})\delta(\omega),\label{eq:s_qe_el}
\end{equation}
respectively, where the magnetic structure factor\cite{Enjalran_2004}
\begin{equation}
\mathbf{F}_{\mu}^{\perp}(\mathbf{Q})=\sum_{i}\mathbf{z}_{i}^{\perp}U_{i\mu}(\mathbf{Q})\exp(\mathrm{i}\mathbf{Q}\cdot\mathbf{r}_{i}).\label{eq:sf}
\end{equation}
We note that we consider only the electronic-spin contribution to the neutron-scattering intensity. There will also be a small contribution from the nuclear spins, of relative magnitude $\sim$$3b_{\mathrm{inc}}^{2}/2C(g_{J}J)^{2}=0.6$\% of the electronic-spin contribution, where $b_\mathrm{inc}=-1.7$\,fm is the nuclear spin-incoherent scattering length for $^{165}$Ho.\cite{Sears_1992} These contributions could in principle be separated using neutron polarization analysis.\cite{Stewart_2009a}

\section{Results and Discussion}\label{sec:results}

\begin{figure}
\begin{center}
\includegraphics[scale=1]{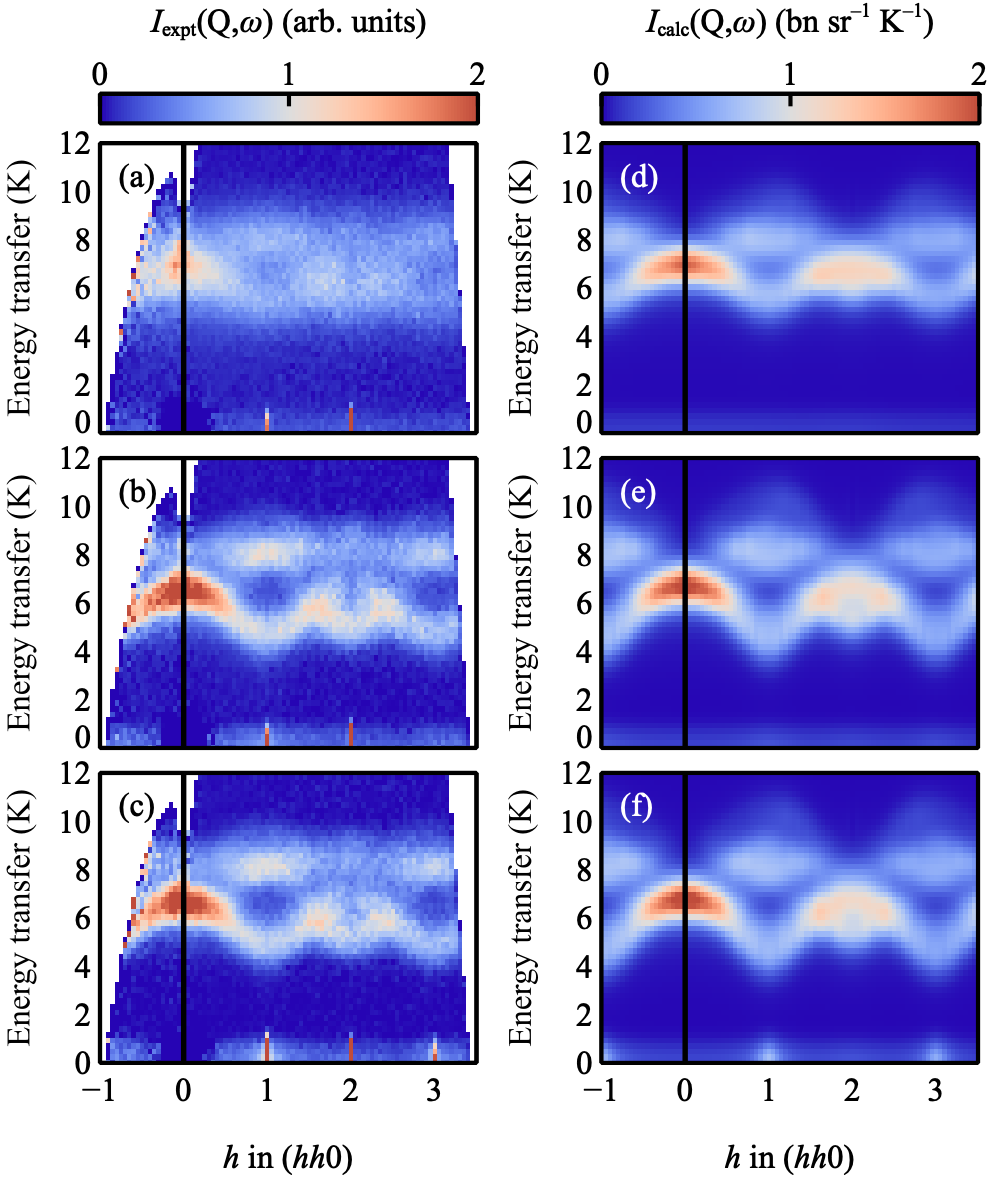}
\end{center}
\caption{\label{fig2}(a) Inelastic neutron-scattering data and model calculations for Ho$_3$Ga$_5$O$_{12}$. Experimental data are reproduced from Ref.~\onlinecite{Zhou_2008}. Data measured at 4\,K, 0.6\,K, and $\sim$0.05\,K are shown in (a), (b), and (c), respectively. Model calculations at 4\,K, 0.6\,K, and 0.35\,K are shown in (d), (e), and (f), respectively. The elastic peaks at $(100)$ and $(200)$ are nuclear Bragg peaks.}\end{figure}

\subsection{Magnetic excitation spectra: model \emph{vs.} experiment}

The previously-published experimental data (from Ref.~\onlinecite{Zhou_2008}) are shown in Fig.~\ref{fig2}(a)--(c). These data were measured using the DCS instrument at the NIST Center for Neutron Research\cite{Copley_2003} at three temperatures, 4\,K, 0.6\,K, and $\sim$0.05\,K. The following features are apparent. At 4\,K, the magnetic scattering consists of weakly dispersive inelastic modes centered at $\omega \sim \Delta$. On cooling the sample to 0.6\,K, two changes occur: the inelastic modes become strongly dispersive with partial softening evident at $(110)$ and $(330)$ reciprocal-space positions, and an elastic diffuse contribution to the scattering also develops. The single-crystal sample considered here undergoes a transition to long-range magnetic order at $T_\mathrm{N}\approx 0.3$\,K.\cite{Zhou_2008} At the base temperature of $\gtrsim0.05$\,K, magnetic Bragg peaks are visible at the $(110)$ and $(330)$ positions, and elastic magnetic diffuse scattering also remains around these same positions. 

To model these observations, we performed calculations of Eqs.~\eqref{eq:s_qe_inel}--\eqref{eq:sf}.
All our calculations were convoluted with the instrumental resolution function, which was approximated as a Gaussian with a FWHM of $1.0$\,K. The experimental data were not normalized on an absolute intensity scale, and so were multiplied by an overall intensity scale factor to match our calculations. Fig.~\ref{fig2}(d)--(f) shows calculations of the magnetic neutron-scattering intensity for the parameter values $\Delta=7.4$\,K, $D = 0.0183$\,K, and $A=0.039$\,K given previously. 
The level of agreement with the experimental data is remarkable given that no fine-tuning of the model parameters has been performed. We also considered the effect of including a nearest-neighbor exchange interaction $\mathcal{K}^\mathrm{nn}$ between local $z$-components of the magnetic moments, but found that this gave qualitatively worse agreement with the data for $|\mathcal{K}^\mathrm{nn}|> 0.1D$, indicating that the interactions of Ho$_3$Ga$_{5}$O$_{12}$ are dominated by $D$.

The approach to magnetic ordering of Ho$_3$Ga$_5$O$_{12}$ is especially informative. The experimental data show that the inelastic soft mode does not soften completely; rather, its minimum energy transfer is approximately $4$\,K both above and below $T_\mathrm{N}$. Accordingly, the magnetic ordering does not occur \emph{via} condensation of the soft mode; instead, the elastic diffuse scattering becomes more intense and eventually sharpens into Bragg peaks on cooling the sample. A possible explanation for this behavior is as follows. In the theory discussed in Section~\ref{sec:model}, magnetic ordering occurs at the highest temperature at which the paramagnetic susceptibility diverges. If the inelastic and elastic contributions to the susceptibility (Eqs.~\eqref{eq:chi_inel} and~\eqref{eq:chi_el}, respectively) are both nonzero, the latter will diverge at the higher temperature, because the term $\lambda_{\mu}(\mathbf{Q})-\lambda$ that involves the magnetic interactions is enhanced by a larger factor $\chi_0^{\mathrm{el}}+\chi_{0}^{\mathrm{in}}(0)$.
Hyperfine interactions can generate a nonzero elastic contribution to the susceptibility, and hence can drive magnetic ordering in two-singlet materials where the electronic interactions alone would be too weak to do so.\cite{Andres_1973,Murao_1971,Murao_1975}  
We will show below that this is the case in Ho$_3$Ga$_5$O$_{12}$.

\subsection{Magnetic ground state}

With a model in hand that explains the experimental data, we now explore its properties. We consider first the low-temperature state below $T_\mathrm{N}$. Experimentally, previous powder neutron-diffraction measurements have shown that Ho$_3$Ga$_5$O$_{12}$ orders with magnetic propagation vector $\mathbf{k}=(0,0,0)$.\cite{Hammann_1977,Hammann_1977a}
The magnetic structure determined by powder neutron diffraction contains six sublattices such that the net moment of the unit cell is zero, as shown in Fig.~\ref{fig1}(c). 
The irreducible representation is $\Gamma_{2+}$ in Miller and Love's notation, and the magnetic space group is $Ia\bar{3}d^{\prime}$. This same magnetic structure is reported in several other several other rare-earth garnets, including Ho$_3$Mn$_x$Ga$_{5-x}$O$_{12}$,\cite{Mukherjee_2017}, Tb$_3$Ga$_{5}$O$_{12}$,\cite{Hammann_1973,Hammann_1977,Hammann_1977a} and Er$_3$Ga$_{5}$O$_{12}$.\cite{Cai_2019}

The predictions of our model correspond well with experiment. First, we obtain the $T_\mathrm{N}$ of our model as the temperature at which the paramagnetic $\chi(\mathbf{q},\omega)$ diverges. Our calculated value of $0.31$\,K  is in good agreement with experimental $T_\mathrm{N}\approx 0.3$\,K for the single crystal studied here. Second, the eigenvector component $U_{i\mu}(\mathbf{k})$ of the lowest-energy mode $\mu$ determines the projection of spin $i$ along its easy axis. The magnetic structure calculated in this way is identical to the one determined from powder neutron diffraction. Our analysis is consistent with the stabilization of this structure by the combination of long-ranged dipolar interactions and easy-axis anisotropy.\cite{Felsteiner_1981}

Our model disagrees with the experimental data in one respect: it predicts a homogeneous magnetic order, which implies that the elastic scattering would contain only magnetic Bragg peaks below $T_\mathrm{N}$. In contrast, the data show both elastic diffuse scattering and magnetic Bragg peaks below $T_\mathrm{N}$. The calculation shown in Fig.~\ref{fig2}(f) was performed just \emph{above}  $T_\mathrm{N}$, at a temperature of 0.35\,K, and shows elastic diffuse scattering at the $(110)$ and $(330)$ positions, indicating the development of static short-range correlations that correspond to finite-sized magnetic domains of the incipient order. While the persistence of diffuse features below $T_\mathrm{N}$ is not explained by our theory, we note that the reported values of $T_\mathrm{N}$ for Ho$_3$Ga$_5$O$_{12}$ vary between $0.15$ and $0.30$\,K, depending on sample and/or measurement protocol.\cite{Hammann_1977,Hammann_1977a,Zhou_2008} It is therefore possible that small variations in stoichiometry could reduce $T_\mathrm{N}$ in regions of the crystal, in agreement with theoretical predictions.\cite{Fulde_1972} Such variations are chemically feasible given that Gd$^{3+}$/Ga$^{3+}$ off-stoichiometry has been reported in isostructural Gd$_3$Ga$_5$O$_{12}$.\cite{Daudin_1982}

\subsection{Relative importance of CEF, frustration, and hyperfine}

We now identify the reason for the suppression of $T_\mathrm{N}$ in Ho$_3$Ga$_5$O$_{12}$. There are two basic scenarios---either $T_\mathrm{N}$ is primarily suppressed by geometrical frustration, or by crystal-field effects---but the hyperfine coupling complicates the picture. To isolate the effects of the various terms in the Hamiltonian, we investigate two further models: one model that removes the hyperfine coupling, and another that removes both the hyperfine coupling and the two-singlet CEF splitting. 

\emph{Model excluding hyperfine coupling} ($A=0$, $\Delta=7.4$\,K, $D=0.0183$\,K).
In the absence of hyperfine coupling, the paramagnetic susceptibility for a two-singlet system is purely inelastic, and Eq.~\eqref{eq:chi_singleion} reduces to\cite{Brout_1966,Jensen_1991}
\begin{equation}
\chi_{0}(\omega)=\frac{2m^{2}\Delta}{\Delta^{2}-\omega^{2}-\mathrm{i}\omega\Gamma}\tanh\left(\frac{\Delta}{2T}\right).\label{eq:chi0_noHF}
\end{equation}
The imaginary part of the interacting susceptibility is given by
\begin{equation}
\mathrm{Im}[\chi_{\mu}(\mathbf{Q},\omega)]=\frac{\Delta^{2}\chi_{0}(0)\omega\Gamma}{[\omega_{\mu}^{2}(\mathbf{Q})-\omega^{2}]^{2}+(\omega\Gamma)^{2}},\label{eq:chi_noHF}
\end{equation}
where the dispersion relation\cite{Brout_1966,Santos_1980}
\begin{equation}
[\omega_{\mu}(\mathbf{Q})]^2 = \Delta^{2}\left\{1-\chi_0(0)[\lambda_{\mu}(\mathbf{Q})-\lambda]\right\}.\label{eq:dispersion}
\end{equation}
Making the assumption that the excitations describe delta-functions in energy, the sum rule [Eq.~\eqref{eq:onsager}] reduces to
\begin{equation}
\frac{\chi_{0}(0)\Delta^{2}}{2NN_{\mathbf{q}}}\sum_{\mu,\mathbf{q}}\frac{\coth[{\omega_{\mu}(\mathbf{q})/2T}]}{\omega_{\mu}(\mathbf{q})}=m^{2}.\label{eq:onsager_rpa}
\end{equation}
The results of Eqs.~\eqref{eq:dispersion}--\eqref{eq:onsager_rpa} were given in Ref.~\onlinecite{Santos_1980} for $N=1$, and are reproduced here for convenience.

The calculated magnetic excitation spectra for $A=0$ resemble those shown in Fig.~\ref{fig2}(d)--(f), except that the elastic contribution to the magnetic scattering becomes zero, as given by Eq.~\eqref{eq:chi_noHF}. However, with $A=0$, it is apparent from Eq.~\eqref{eq:dispersion} that magnetic ordering can only occur if the interactions are sufficiently strong, so that $R_\mathrm{c} = (2m^2/\Delta)\mathrm{max}[\lambda_{\mu}(\mathbf{Q})-\lambda]\geq1$ at $T=0$. This is not the case for Ho$_3$Ga$_5$O$_{12}$, in which $R_\mathrm{c}=0.71$ at $T=0$. Therefore, long-range magnetic ordering in Ho$_3$Ga$_5$O$_{12}$ is enabled by the hyperfine interaction; if this were absent, Ho$_3$Ga$_5$O$_{12}$ would remain a correlated paramagnet to zero temperature. Hyperfine-assisted magnetic ordering is uncommon, and its $T_\mathrm{N}$ is relatively high in Ho$_3$Ga$_5$O$_{12}$ compared to other known examples\cite{Nicklow_1985} because the hyperfine coupling is strong and $R_\mathrm{c}$ is not too much smaller than unity.

\emph{Model excluding hyperfine coupling and crystal-field splitting} ($A=0$, $\Delta=0$, $D=0.0183$\,K):
In the absence of both hyperfine coupling and two-singlet splitting, the paramagnetic susceptibility is (quasi-)elastic, and Eq.~\eqref{eq:chi_singleion} reduces to a Curie law,
\begin{equation}
\chi_{0}(\omega) = \delta_{\omega0} \frac{m^{2}}{T}.\label{eq:chi0_noHF_noSI}
\end{equation}
The sum rule [Eq.~\eqref{eq:onsager}] reduces to
\begin{equation}
m^{2} = \frac{T}{NN_{\mathbf{q}}}\sum_{\mu,\mathbf{q}} \chi^{\mathrm{el}}_{\mu}(\mathbf{q}).\label{eq:onsager_noHF_noSI}
\end{equation}

We find that this model undergoes a transition to long-range magnetic order at a temperature of 1.74\,K, which is comparable to the strength of the dipolar interaction. This result demonstrates that the suppression of magnetic ordering observed in Ho$_3$Ga$_5$O$_{12}$ is not a consequence of geometrical frustration, as was suggested previously.\cite{Zhou_2008,Mukherjee_2017} Instead, it is a consequence of single-ion quantum fluctuations induced by the two-singlet crystal-field ground state. The reason why Ho$_3$Ga$_5$O$_{12}$ is not geometrically frustrated can be understood by considering the dipolar interaction, Eq.~\eqref{eq:dipolar}. The bond-dependence of this interaction is such that, for each Ho$^{3+}$ ion, Eq.~\eqref{eq:dipolar} is positive for two of its nearest neighbors and negative for the other two. The energy of the nearest-neighbor dipolar interaction is  hence minimized by the ordered magnetic structure shown in Fig.~\ref{fig1}(c), in which each spin has two neighbors parallel to their quantization axes and two neighbors antiparallel to them.

\subsection{Comparison of reaction-field and RPA results}

\begin{figure}
\begin{center}
\includegraphics{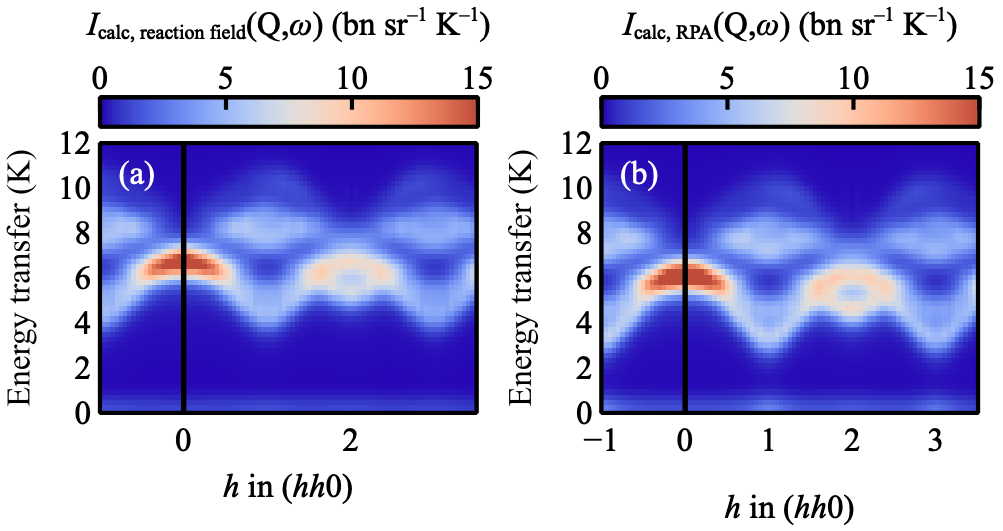}
\end{center}
\caption{\label{fig3}Calculations of the magnetic excitation spectrum of Ho$_3$Ga$_5$O$_{12}$ at $T=1.0$\,K using (a) the reaction-field approximation and (b) the RPA.}\end{figure}

Finally, we compare results obtained using the reaction-field approximation with the standard RPA. This is achieved by setting the reaction-field parameter $\lambda$ to zero in Eqs.~\eqref{eq:chi_inel} and \eqref{eq:chi_el}. Figs.~\ref{fig3}(a) and \ref{fig3}(b) show calculations of the magnetic excitation spectra using the reaction-field approximation and the standard RPA, respectively. Both calculations are performed at $T=1$\,K. The RPA-only calculation overestimates the bandwidth and degree of softening of the magnetic excitations. 
Moreover, the magnetic ordering temperature of $0.66$\,K obtained from standard RPA is too large by a factor of two. These results demonstrate that the reaction-field approximation represents a significant improvement over the standard RPA, and therefore provides a straightforward alternative to other extensions of mean-field theory.\cite{Jensen_1994,Jensen_2011} These favorable results are consistent with previous studies of the reaction-field approximation for various models, including geometrically-frustrated examples.\cite{Santos_1980,Logan_1995,Wysin_2000,Wysocki_2008,Hohlwein_2003,Conlon_2010}

\section{Conclusions and Outlook}\label{sec:conclusions} 

Our modeling study of inelastic neutron-scattering data\cite{Zhou_2008} unambiguously determines the magnetic Hamiltonian of Ho$_3$Ga$_5$O$_{12}$ in terms of the CEF, dipolar interactions, and nuclear hyperfine coupling. We find that the suppression of long-range magnetic ordering is due not to geometrical frustration, as was suggested previously,\cite{Zhou_2008,Mukherjee_2017} but instead to the intrinsic transverse field generated by its two-singlet CEF ground state. The eventual magnetic ordering in Ho$_3$Ga$_5$O$_{12}$ is driven by the nuclear hyperfine coupling, in qualitative agreement with early theoretical calculations.\cite{Hammann_1973} Our results raise the intriguing possibility of tuning the value of $\Delta$ by chemical substitutions of Ga$^{3+}$ for Al$^{3+}$, Sc$^{3+}$, or In$^{3+}$.\cite{Mukherjee_2017b}
Moreover, the effect of doping Ho$^{3+}$-based garnets\cite{Mukherjee_2017a} may allow theories of impurity effects in two-singlet systems\cite{Fulde_1972} to be tested.

Our results benchmark theoretical predictions of the excitation spectra of interacting two-singlet systems against modern neutron scattering data. This comparison has been a longstanding goal of condensed-matter physics, and has been pursued in several systems, including PrX$_2$Si$_2$ ($\textrm{X}=\textrm{Fe,\,Ru}$),\cite{Blaise_1995,Mulders_1997} HoF$_3$,\cite{Leask_1994,Jensen_1995} LiTb$_x$Y$_{1-x}$F$_4$,\cite{Lloyd_1990} as well as variants such as three-singlet\cite{Kawarazaki_1995} and singlet-triplet\cite{Birgeneau_1971,Christianson_2007} systems. Our results significantly advance previous studies by modeling both the energies and the intensities of the dispersive paramagnetic excitations over a wide range of reciprocal space. It is especially interesting to compare our results with inelastic neutron-scattering experiments on the Ising doublet system LiHoF$_4$ in an applied transverse magnetic field.\cite{Ronnow_2005} As the value of this applied field is increased, mode softening is observed; however, this softening remains partial and is arrested by the onset of long-range magnetic order due to the nuclear hyperfine coupling.\cite{Ronnow_2005} These observations qualitatively resemble our results, demonstrating that the intrinsic transverse field generated by the CEF in Ho$_3$Ga$_5$O$_{12}$ has an equivalent effect to an externally-applied transverse field in LiHoF$_4$.

From a methodological perspective, we have shown that the reaction-field approach\cite{Santos_1980}---which combines the RPA\cite{Brout_1966} with the total-moment sum rule---can accurately model magnetic excitation spectra in systems where CEF levels and magnetic interactions have similar energies. It represents a significant improvement over the standard RPA while retaining its low computational cost and general applicability. Because the reaction-field approach has already been shown to yield accurate results in frustrated systems in the absence of CEF effects,\cite{Hohlwein_2003,Conlon_2010} we are optimistic that it will now prove useful for topical frustrated systems in which CEF effects are important, such as Pr$^{3+}$-based pyrochlores,\cite{Zhou_2008a,Wen_2017,Martin_2017} osmate pyrochlores,\cite{Zhao_2016} spin-chain SrHo$_2$O$_4$,\cite{Young_2013} spinel NiRh$_2$O$_4$,\cite{Chamorro_2018} and tripod kagome systems.\cite{Dun_2016,Dun_2017,Paddison_2016, Ding_2018} 

Finally, our study demonstrates the essential role played by hyperfine coupling in systems with singlet CEF ground states.\cite{Triplett_1973,Andres_1973} The hyperfine coupling has typically been neglected in current models of spin-liquid candidates, but our results suggest that this assumption may often be invalid in non-Kramers systems.\cite{Wen_2017,Duijn_2017,Zhao_2016,Ding_2018} In such materials, singlet ground states are guaranteed if the point symmetry is sufficiently low; however, even if this condition is not met and a doublet ground state is expected, singlet ground states may nevertheless arise from structural disorder\cite{Wen_2017,Duijn_2017} or from perturbations associated with the measurement probe.\cite{Foronda_2015} Although the hyperfine coupling is largest for Ho$^{3+}$ ($I=7/2$), as considered here, we anticipate that the effects will also be significant for Pr$^{3+}$ frustrated magnets ($I=5/2$),\cite{Triplett_1973} providing an important avenue for future research on these interesting materials.\cite{Zhou_2008a,Wen_2017,Martin_2017}

\section*{Appendix A}
The single-ion Hamiltonian projected in the space of two singlet states can be written as
\begin{align}
\mathcal{H} = \dfrac{\Delta}{2}\mathbf{1}\otimes{\sigma}^{x} + A m {I}^{z}\otimes{\sigma}^{z}
\end{align}
where ${\sigma}^{\alpha}$ are $2\times 2$ unit Pauli matrices, ${I}^{\alpha}$ is the angular momentum operator of nuclear spin of dimension $(2I+1)\times(2I+1)$, $m = \left|\left\langle 0 | {J}^{z}| 1\right\rangle\right|$ is the matrix element between the two singlet states, and $A$ is the nuclear hyperfine coupling strength. This Hamiltonian commutes with the ${I}^{z}$ operator, so we can work in the subspace with different eigenvalues $I^{z}$, where the Hamiltonian is reduced to a $2\times2$ matrix
\begin{align}
\mathcal{H} = h^{x}{\sigma}^{x} + h^{z}{\sigma}^{z},
\end{align} 
where $h^{x} = \Delta/2$ is a transverse field and $h^{z} = A m I^{z}$ is a longitudinal field. It is straightforward to diagonalize the Hamiltonian and obtain energies of the ground state and the excited state,
\begin{align}
E_{i}(I^{z}) = -(-1)^{i}\sqrt{(h^{x})^{2}+(h^{z})^{2}}\,,\quad i = 0,1,
\end{align}
and the transition matrix elements
\begin{align}
|\left\langle I^{z}, 0 | {\sigma}^{z}| I^{z}, 0\right\rangle| & = |\left\langle I^{z}, 1 | {\sigma}^{z}| I^{z}, 1\right\rangle| \\
& = \frac{(h^{z})^2}{(h^{x})^2+(h^{z})^2}, \\ 
|\left\langle I^{z}, 0 | {\sigma}^{z}| I^{z}, 1\right\rangle| &=  |\left\langle I^{z}, 1 | {\sigma}^{z}| I^{z}, 0\right\rangle| \\
&=\frac{(h^{x})^2}{(h^{x})^2+(h^{z})^2}.
\end{align}
There are no transitions among states with different $I^z$.

\section*{Acknowledgements}

J.A.M.P. acknowledges financial support from Churchill College, Cambridge (UK), and an ILL Visiting Scientist position during which part of this work was completed. The authors acknowledge Bruno Tomasello and Tim Ziman for helpful discussions. The work of X.B., Z.L.D. and M.M. at Georgia Tech (single-ion Hamiltonians of non-Kramers magnets) was supported by the U.S. Department of Energy, Office of Basic Energy Sciences, Materials Sciences and Engineering Division under award number DE-SC-0018660. C.R.W. acknowledges funding from NSERC (Discovery Grant and CRC program).

\end{document}